\documentclass{rQUF2e}
\usepackage{graphics}
\usepackage{epsfig}

\makeatletter
\@addtoreset{chapter}{part}
\makeatother

\newcommand\lsim{\mathrel{\rlap{\lower4pt\hbox{\hskip1pt$\sim$}}
    \raise1pt\hbox{$<$}}}
\newcommand\gsim{\mathrel{\rlap{\lower4pt\hbox{\hskip1pt$\sim$}}
    \raise1pt\hbox{$>$}}}
    
\newcommand\be{\begin{equation}}
\newcommand\bea{\begin{eqnarray} \nonumber }
\newcommand\ee{\end{equation}}
\newcommand\eea{\end{eqnarray}}

\begin{document}

\title{Risk Premia: \\
Asymmetric Tail Risks and Excess Returns}

\author{Y. Lemp\'eri\`ere, C. Deremble, T. T. Nguyen\\
P. Seager, M. Potters, J. P. Bouchaud \\
Capital Fund Management, \\ 
23 rue de l'Universit\'e, 75007 Paris, France \\
}
\maketitle

\begin{abstract}
We present extensive evidence that ``risk premium'' is strongly correlated with tail-risk skewness but very little with volatility. 
We introduce a new, intuitive definition of skewness and elicit an approximately linear relation between the Sharpe ratio of various risk premium strategies 
(Equity,  Fama-French, FX Carry, Short Vol, Bonds, Credit) and their negative skewness. 
We find a clear exception to this rule: trend following has both positive skewness and positive
excess returns. This is also true, albeit less markedly, of the Fama-French ``Value'' factor and of the ``Low Volatility'' strategy. 
This suggests that some strategies are not risk premia but genuine market anomalies. Based on our results, we propose an objective criterion to 
assess the quality of a risk-premium portfolio.

\end{abstract}
\section{Introduction}

One of the pillars of modern finance theory is the concept of ``risk premium'', i.e. that a more risky investment should, 
on the long run, also be more profitable. The argument leading to such a conclusion looks, at least at first sight, rock solid: given the choice between 
two assets with the same expected return, any rational investor should prefer the less risky one. The resulting smaller demand for riskier assets should 
make their prices drop, and therefore their yields go up, until they become attractive enough to lure in investors in spite of their higher risk. This simple
idea appears to explain why stock indices on the long run yield higher returns than monetary funds, for example. It is also at the heart of the well-known Capital 
Asset Pricing Model (CAPM), which predicts that the excess return of a given stock (over the risk free rate) is proportional to its so-called ``$\beta$'', i.e. to its covariance 
with the market risk \citep{CAPM}. 

The idea has become so pervasive that its logic has actually inverted: efficient market theorists assume that any return in excess
of the risk free rate should in fact be associated with a non-diversifiable risk factor of some sort -- sometimes argued to be unobservable or ``latent''. 
It is often invoked to do away with any market anomaly or discrepancies between market prices and theoretical prices, 
as it allows one to introduce an extra fitting parameter conveniently called the ``market price of risk''.
\footnote{In the context of option pricing, see e.g. \citep{Fouque}. 
Note, however, that since the market price of risk itself is usually found to be strongly fluctuating in time, 
the predictive strength of these theories becomes frustratingly close to zero.}

Although the argument underlying the existence of risk premia is cogent, estimating and rationalising their order of magnitude in financial markets is still very much a matter of debate. 
This is quite surprising in view of the importance of the issue for the asset management industry as a whole, and reveals how primitive is our
understanding of financial prices and returns. Empirically, the equity risk premium is found to be in the range $3 - 9 \%$ 
depending on the period and country \citep[see][and our own data in Sect. II below]{MP2,ERP,Fernandez}. 
Mehra \& Prescott have pointed out that these values are too high to be explained within a general equilibrium model of the economy and have 
coined this discrepancy the ``equity premium puzzle'' \citep{MP}.\footnote{However, one should note that the Mehra \& Prescott framework fails to explain the high level of the volatility of the
stock market, i.e. it does not explain the ``excess volatility puzzle'' in the first place!} More recently, several authors \citep{Ang,Blitz,Ang2,LowVol,BAB,LowVol_us} have reported an 
{\it inverted} relation between the volatility
of a stock (or its $\beta$) and its excess return. In other words, contrarily to intuition, less volatile stocks appear to be more profitable.

As pointed out in numerous papers -- many of them quite recent -- the problem may reside in the very definition of risk. Classical theories identify risk with 
volatility $\sigma$. This (partly) comes from the standard assumption of a Gaussian distribution for asset returns, which is entirely characterised by its first
two moments: mean $\mu$ and variance $\sigma^2$. But in fact fluctuations are known to be strongly non Gaussian, and investors are arguably not much concerned by small fluctuations around
the mean. Rather, they fear large negative drops of their wealth, induced by rare, but plausible crashes. These negative events are not captured by the 
r.m.s. $\sigma$ but rather contribute to the negative skewness of the distribution. Therefore, an alternative idea that has progressively emerged in the 
literature is that a large contribution to the ``risk premium'' is in fact a compensation for holding an asset that provides positive average returns but may
occasionally erase a large fraction of the accumulated gains. This was in fact proposed as a resolution of the Mehra \& Prescott ``equity premium puzzle'' by 
\citet{Kraus,Rietz} and then in many follow up papers with a similar message, e.g. \citep{HarveySiddique,Barro,SantaClara,Bollersev}, 
 \citep[see also][]{Kelly,Others1,Others2}. Risk premium might be more a question of skewness and negative tail events than related to volatility {\it per se}. 

A more formal way to frame this story is in terms of the utility function supposed to describe investors' preferences. 
Assuming that the ``happiness'' of the investor depends on his/her wealth $W$ as a certain function $U(W)$, the preference for higher profits imposes that $U(W)$ 
is a monotonously increasing function of $W$, whereas the condition $U''(W) \leq 0, \forall W$ ensures that a less volatile investment with the same expected gain is always preferred. 
Assuming that the proposed investment yields an uncertain final wealth $W$ distributed around a mean $\mathbb{E}[W]$ with small fluctuations of variance $\sigma^2$, 
and expanding in powers of the difference $W - \mathbb{E}[W]$, one finds that the expected utility is given by:\footnote{
A way to avoid this expansion is to assume an exponential (CARA) utility function $U(W) \equiv - \exp(-\lambda W)$ and Gaussian fluctuations for $W$. In this case,
$$
\mathbb{E}[U(W)] \equiv -\exp [ - \lambda \mu  + \frac12 \lambda^2 \sigma^2].
$$}
\be\label{risk-reward}
\mathbb{E}[U(W)] = U(\mathbb{E}[W]) - \frac12 \lambda^2 \sigma^2 + ...\qquad \lambda^2 := -U''(\mathbb{E}[W]) \geq 0.
\ee
Hence, within a small risk expansion, utility reduction is entirely described by the variance $\sigma^2$. The above approximate formula is the starting point to 
most risk-return portfolio optimisation schemes in the classical literature. 

But as intuition suggests, skewness matters. Lottery experiments \citep{Lotteries1,Lotteries2} demonstrate that agents prefer lotteries with very large potential gains -- 
even when the expected gain is negative -- to lotteries with small positive expected gain but no possibility of large payoffs. 
Symmetrically, one expects agents to shy away from investments with a high degree of negative skewness. Clearly, this cannot be captured by Eq. (\ref{risk-reward}) above where skewness is absent; it
may well be the case that the whole concept of utility is anyway insufficient to account for behavioral biases, see \citep{Kahneman}. 
Still, a large number of papers have therefore been devoted to generalising Markowitz's portfolio optimisation and the CAPM to account for skewness preferences \citep{Kraus,HarveySiddique,CAPMskew,Harvey}. 
In terms of utility functions, one can show that the additional condition $U'''(W) \geq 0, \forall W$ ensures that more positively 
skewed investments are preferred \citep{Uskew}.
\footnote{More precisely, this preference ordering requires in general the distributions to be ``skewness comparable'', which means that their cumulative functions must 
cross twice and only twice \citep{Oja,Uskew}. This turns out to be often the case in financial applications, see below and Appendix.}

Our work is clearly in the wake of the above mentioned literature on skewness preferences and tail-risk aversion. We will present extensive evidence that ``risk premium'' is indeed 
strongly correlated with the skewness of a strategy but very little with its volatility, not only in the equity world -- as was emphasised by previous authors -- but in other sectors as well. 
We will investigate in detail many classical so-called ``risk premium'' strategies (in equities, bonds, currencies, options and credit) and 
elicit a linear relation between the Sharpe ratio of these strategies and their negative skewness. 
We will find however that some well-known strategies, such as trend following and to a lesser extent the Fama-French ``High minus Low'' factor and the ``Low Vol'' strategy, 
are clearly not following this rule, suggesting that these strategies are not risk premia but genuine market anomalies.

Compared to the previous abundant literature, the present results are new in different respects. First, at variance with most previous investigations (that mostly focusses on stock markets), 
we do not attempt to frame our empirical analysis within the constraining framework of asset pricing and portfolio theory, but rather let the data speak for itself. This is specially important when studying, as we do here, 
risk premia across a much larger universe of assets, where the notion of a global ``risk factor'' (generalizing the market factor in the equity space) is far from clear, 
as we discuss in Sect. 8 below.  Second, we introduce a simple way to plot the returns of a portfolio that reveals its skewness to the ``naked eye'' and suggests an intuitive and robust definition of skewness that is 
much less sensitive to extreme events. Third, our empirical conclusion that for a wide spectrum of ``risk premia'' strategies, skewness rather than volatility is a determinant of returns is, to the best of 
our knowledge, new, as is the finding that some investment strategies -- like trend following -- seem to behave quite differently. 

We first start in Sect. 2 with the equity market as a whole and revisit the equity risk premium world-wide, and its (negative) correlation with the volatility.
We then introduce our new, intuitive definition of skewness that we use throughout the paper and that we justify in the Appendix.  
We focus on the Fama-French factors in Sect. 3 and study the statistics of market neutral portfolios, including a ``Low Volatility'' portfolio. We move on to the fixed income world (Sect. 4), 
where we again build neutral portfolios. 
Sect. 5 is devoted to an account of risk premia on currencies (the so-called ``Carry Trade''), and finally, in Sect. 6, to the paradigmatic case of selling options. We summarise our findings in Sect. 7 with a 
suggestive linear relation between the Sharpe ratio and the skewness of all the Risk Premium strategies investigated in the paper, and discuss some exceptions to the rule 
-- i.e. positive Sharpe strategies with zero or positive skewness -- that we define as ``pure $\alpha$'' strategies. We then end the paper in Sect. 8 by comparing our results to the extended CAPM 
framework {\it \`a la} \citet{Kraus,HarveySiddique}. We propose instead a simple argument to rationalise for the Sharpe ratio/skewness trade-off and an objective criterion to assess the quality of a risk premium portfolio. 

\section{Back to Basics: Risk Premium in Global Equity Markets, and a new definition of Skewness}

\subsection{No ``volatility premium'' in stock indices}

According to the standard framework described in the previous section, investing in the stock market rather than in risk-free instruments only makes sense if
stocks provide, in the long run, better returns than the risk-free rate. The existence and the strength of this equity risk premium (ERP) 
has been much debated amongst both the academic community and the asset management industry. The general consensus (confirmed below) is that this premium exists 
and is in the range $3 - 9 \%$, with large fluctuations, both across epochs and countries \citep{MP2,ERP,Fernandez}. Within a classical general equilibrium framework, 
this premium appears to be too large. With such a level of ERP, rational investors should put money more massively in equities, 
and the demand for bonds should be much weaker than what we observe, which should lead to a smaller risk premium. 
This is the ``equity premium puzzle'' brought to the fore by \citet{MP}.

\begin{figure}
\begin{center}
\epsfig{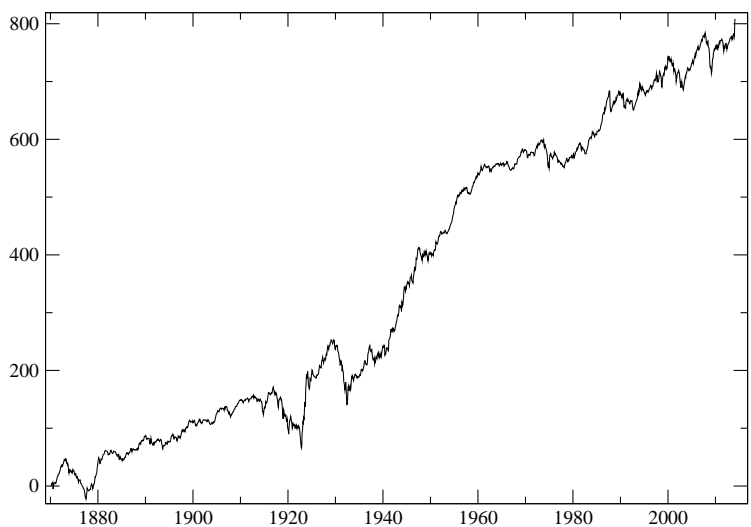}
\caption{Cumulated ERP since 1870, corresponding to a strategy where one holds a long portfolio of indices, with equal weight on all
available contracts at any instant of time, i.e. $1$ in 1870 and $27$ in 2014. The t-stat is $4.2$.}
\label{pnlERP}
\end{center}
\end{figure}

In order to estimate the ERP, it is customary to compute the total return of indices (dividend included), and to subtract the risk-free rate. \footnote{This is virtually equivalent to buying a future contract on indices.} 
We have therefore computed this cumulated difference for various countries, using the 10Y government rate as a risk-free asset 
(using 3-month bills would only give a more flattering result, since short rates are usually lower than 10Y rates). 
The result is, as expected, overwhelmingly in favour of the existence of a risk premium, as can be seen from Table \ref{ERP} and Fig. \ref{pnlERP}. 
All countries but one display positive excess return on the long run (some time series go back as far as 1870). The global t-stat of a strategy equi-invested in all available indices 
at any moment in time is $4.2$. The average value (weighted by the number of available years for each country) is found to be  $\approx 5 \%$ with a cross-sectional dispersion of $\pm 3\%$, 
compatible with the numbers quoted above \citep{MP2,ERP,Fernandez}. 

\begin{figure}
\begin{center}
\epsfig{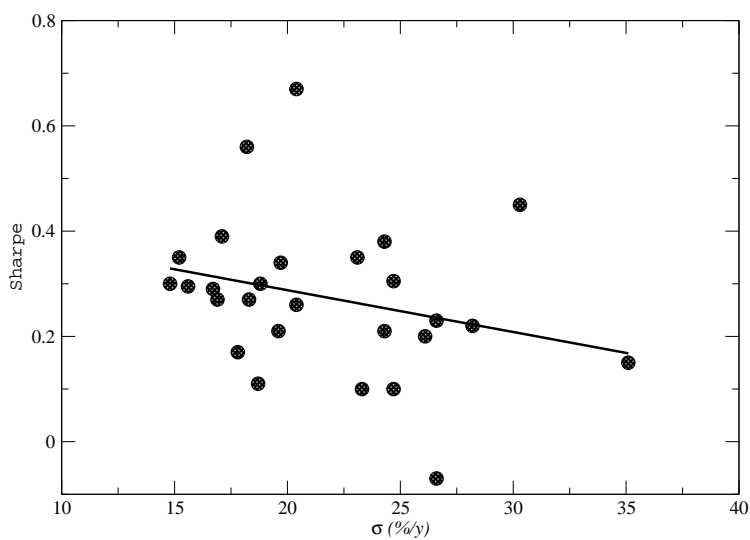}
\caption{Sharpe ratio of the excess return (over the 10Y bond) for each index listed in Table \protect{\ref{pnlERP}} as a function of the volatility of that index. We also plot the
regression line, with a {\it negative} slope and a correlation coefficient $\rho \approx -0.27$. Note that the (flat) average Sharpe ratio over all contracts is ${\overline{\cal S}} \approx 0.3$.}
\label{Data}
\end{center}
\end{figure}

In the fourth column of Table \ref{ERP}, we give the average annual volatility of the corresponding index. This allows us to compute the Sharpe ratio ${\cal S}=\mu/\sigma$ of 
each risk premium. Note indeed that the value of the risk premium itself is not meaningful since it can be arbitrarily leveraged. 
Therefore here and throughout the paper we use the a-dimensional Sharpe ratio to meaningfully compare different risk premia. 
Fig. \ref{Data} shows a scatter plot of ${\cal S}$ as a function of the volatility of the index, together with a regression line. 
We see that the two quantities are in fact {\it negatively} correlated ($\rho \approx -0.27$). Even the risk premium itself (i.e. without dividing by the volatility $\sigma$) 
is found to be weakly anti-correlated with $\sigma$ ($\rho \approx -0.1$) but this is not statistically significant. This is at odds with the standard interpretation of the risk premium, but 
confirms once again the ``low-volatility'' anomaly reported by many authors \citep{Ang,Blitz,Ang2,LowVol,BAB,LowVol_us}: the risk-adjusted returns of low-vol stocks is found to be higher than 
for high-vol stocks.

\begin{table}
\begin{center}
\begin{tabular}{|c|c|c|c|c|c|c|c|}
\hline
\hline
Country & Start date & Premium & Vol. & Sharpe & Skew $\zeta_3$ &  Skew $\zeta^*$ & Co-skew/US \\
\hline
Australia & 1983 & 4.77 & 16.7 & 0.29 & -2.0 &-2.35 & 0.03\\
Canada & 1934 & 4.44 & 14.8 & 0.30 & -0.6 &-1.68 & 0.03\\
Germany & 1870 & 4.23 & 19.6 & 0.295 & 1.7 &-0.31 & 0.05 \\
France & 1898 & 6.65 & 19.7 & 0.21 & 2.7 & 1.05 & 0.03 \\
Finland & 1970 & 8.09 & 23.1 & 0.35& 0.2 & 0.8 & 0.01\\
Italy & 1925 & 6.24 & 26.6 & 0.34 & 1.7 & 1.30 & 0.01\\
Hong Kong & 1996 & 5.24 & 26.1 & 0.23 & 0. &  -1.87 & 0.01\\
Hungary & 1999 & 2.51 & 24.7 & 0.20 & -0.2 &-1.79 & -0.04\\
India & 1988 & 13.6 & 30.3 & 0.10 & 0.5 & 0.63 & -0.17\\
Indonesia & 2009 & 13.8 & 20.4 & 0.67 & 0.4 & -3.26 & 0.01\\
Japan & 1980 & 2.04 & 18.7 & 0.45 & -0.1 &-0.85 & 0.12 \\
South Korea & 2000 & 9.33 & 24.3 & 0.11 & -0.2 & -0.08 & -0.02\\
Malaysia & 1974 & 6.32 & 28.2 & 0.38 & 0.1 &-0.14 & -0.09\\
Mexico & 2001 & 10.1 & 18.2 & 0.56 & -0.5 &-2.26 & 0.12\\
The Netherlands & 1957 & 4.57 & 16.9 & 0.22 & -0.5 & -1.7 & -0.03\\
New Zealand & 1984 & 3.02 & 17.8 & 0.27 & -0.1 & -1.67 & 0.05 \\
Norway & 1969 & 5.21 & 24.3 & 0.21 & -0.5 & 0.17 & 0.05\\
Philippines & 1996 & -1.96 & 26.6 & 0.17 & 0.2 & -0.42 & 0.02\\
Poland & 1999 & 2.29 & 23.3 & -0.07 & 0.1 & -0.62 & 0.05\\
Singapore & 1973 & 7.51 & 24.7 & 0.10 & -0.2 & -0.86 & -0.05\\
South Africa & 1983 & 5.26 & 20.4 & 0.27 & -0.6 & -0.14 & 0.03 \\
Spain & 1948 & 5.25 & 18.3 & 0.39 & -0.1 & -0.12 & 0.05\\
Sweden & 1960 & 6.69 & 17.1 & 0.305 & -0.1 & -0.42 & -0.05\\
Switzerland & 1966 & 4.58 & 15.6 & 0.26 & -0.5 & -2.05 & 0.07 \\
Thailand & 1997 & 5.25 & 35.1 & 0.15 & 0.1 & -0.94 & 0.0 \\
United Kingdom & 1958 & 5.71 & 18.8 & 0.30 & 1. & -2.31 & 0.06\\
United States & 1871 & 5.33 & 15.2 & 0.35 & 0.3 & -0.32 & 0.06\\
\hline
\hline
\end{tabular}
\caption{Average annualized excess return (``risk premium'') in \%, annualized volatility in \%, Sharpe ratio, standard skewness $\zeta_3$ and ranked P\&L skewness $\zeta^*$ for various 
indices over the 10Y bond rates, all based on monthly returns. The average Sharpe ratio is $\approx 0.3$. Start dates as indicated in the second column; end date Jan 2014. 
Note that while the ranked P\&L skewness $\zeta^*$ is negative across most markets as expected, the classical third cumulant skewness $\zeta_3$ is very noisy and has a random sign. 
In the last column, we give the co-skewness with the US index, in the spirit of \citet{Kraus,HarveySiddique}.
Data from Global Financial Data, ({\tt www.globalfinancialdata.com}).}
\label{ERP}
\end{center}
\end{table}

\subsection{Downside tail risk and skewness premium}

\begin{figure}
\begin{center}
\epsfig{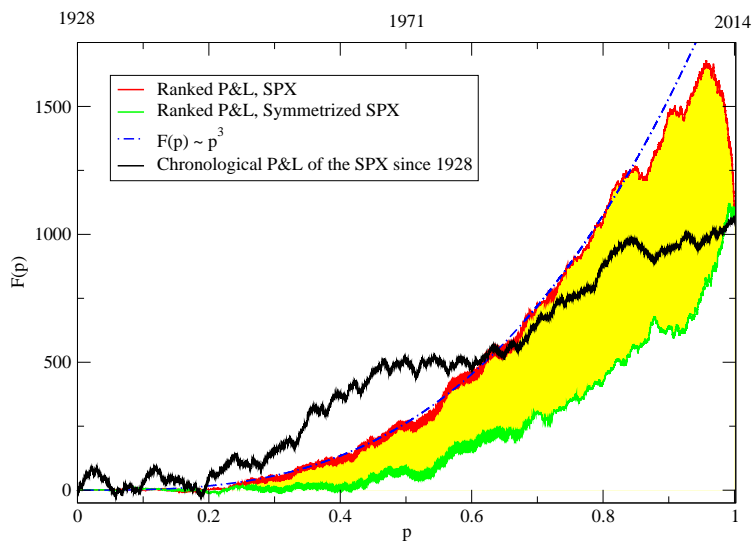}
\caption{The ranked amplitude P\&L representation: plot of the cumulated daily P\&L $F(p)$ for the SPX (at constant risk) since 1928, 
 as a function of the normalised rank of the {\it amplitude} of the 
returns, $p=k/N$ (in red). The standard, chronological P\&L corresponding to holding the US equity market index at constant risk is shown in black, 
that by construction ends at the exact same point. We also show for comparison (in green) $F_s(p)$, corresponding to returns symmetries around the same global mean 
(i.e. $F_s(p=1)=F(1)$, again by construction). 
Our definition of the skewness $\zeta^*$ is, up to a negligible correction, minus the area enclosed between $F(p)$ and $F_s(p)$ (shown in yellow). }
\label{skewSPX}
\end{center}
\end{figure}

However, something interesting occurs when we suggestively re-plot the cumulated ERP (say for the risk managed SPX since 1928 \footnote{Here and below, we will consider 
``risk managed'' strategies that run at approximately constant risk over time. This is achieved by normalising the position by a twenty-day exponential moving average of 
the volatility of the bare strategy itself. In the case of the SPX, this simply amounts to the recent volatility of the SPX itself.}) in the following way. 
Instead of considering the returns of the strategy in chronological order, we sort these returns in terms of their absolute value and plot the 
cumulated ERP $F(p)$ as a function of the normalised rank $p=k/N$, starting from the return with the smallest amplitude ($k=1$) and ending with the largest ($k=N$). 
The result in the case of the US since 1928 is shown in Fig. \ref{skewSPX}. We immediately see that while the small returns contribute positively to the 
average, the largest returns, contrarily, lead to a violent drop of the P\&L. Strikingly, the $5 \%$ largest returns wipe out half of the ERP of the
$95 \%$ small-to-moderate returns.

In the same graph, we also show the P\&L $F_s(p)$ that would have been observed if the distribution of returns was exactly symmetrical around the same mean -- 
i.e. such that the final point $F_s(p=1)$ coincides with $F(p=1)$.\footnote{Technically, this amounts to transforming 
the returns $r_t$ as: $m + \epsilon_t (r_t -m)$ where $m$ is the average return and $\epsilon_t = \pm 1$ an independent random sign for each $t$.} 
In this case, it is easy to show that the P\&L is (for large $N$) a monotonously increasing function of the normalised rank $p$. One finds that generically $F(p)$ behaves as $p^3$ at small $p$ --
explaining the strong curvature seen in Fig. \ref{skewSPX}. The comparison between real returns and symmetries returns therefore reveals the {\it strongly skewed} nature 
of the ERP. 

The ranked amplitude P\&L representation in fact suggests a new general definition of skewness, as follows: one 
first normalises the returns such that their average is zero and their rms is unity. Then the ranked amplitude P\&L is plotted, and defines a certain function $F_0(p)$, such that
$F_0(0)=F_0(1)=0$. By construction, the symmetries version $F_{0s}(p)$ is now zero (up to small fluctuations). Our definition of the skewness $\zeta^*$ is simply (minus) the area underneath the 
function $F_0(p)$:\footnote{One could actually directly take the area between $F(p)$ and $F_s(p)$ as our definition of skewness, i.e. without first removing the average drift $\mu$. Up to relative corrections of order
$(\mu/\sigma)^2$, the result would be identical. For all purposes, this correction is extremely small, at least for daily returns (i.e. less than $10^{-3}$).}
\be
\zeta^* := - 100 \int_0^1 {\rm d}p \, F_0(p),
\ee
where the arbitrary factor $100$ is introduced such that the skewness is of order unity.

We give several properties of $F_0(p)$ in the Appendix and show how $\zeta^*$ can be related to the classical definition of skewness $\zeta_3$ defined as the third cumulant of the distribution of returns. 
Although standard, this definition is awkward for heavy-tailed distributions (as those describing financial returns) since a few extreme events can completely dominate the empirical determination of the
skewness $\zeta_3$. In these cases, it is strongly advised to use low-moment estimates of the skewness, such as the normalized mean-minus-median for example. We prefer our new definition $\zeta^*$ above, 
not only because of its intuitive interpretation, but also because $\zeta^*$ can be defined even when the mean of the distribution diverges. [Note however that all the results reported in this paper are actually robust 
to the chosen definition of skewness, provided it is based on low moments.] We also show in the Appendix that whenever $F_0(p)$ has a {\it single hump}, as is usually
the case in financial applications, then the cumulative distribution of returns crosses its symmetries version {\it twice and only twice}. This means, as announced in the introduction, that investors with a utility $U$ 
such that $U'''> 0$ will shy away from investments with negative $\zeta^*$, i.e. will demand a higher risk premium. 

In the case of the US market factor, we find $\zeta^* \approx -1.47$ for daily returns (for a Sharpe ratio of $0.57$), and $\zeta^*_{\text{month}} \approx -0.32$ for monthly returns (while the 
standard skewness $\zeta_{3,\text{month}} \approx +0.3$ is actually positive!). When analysing the data given in Table 1, we do find a {\it positive} correlation between $-\zeta^*$ 
and the ERP across different countries, $\rho \approx 0.2$. This is encouraging but not particularly convincing.\footnote{Note that the correlation between the ERP and the standard skewness $\zeta_3$ is 
close to zero, due to the large amount of measurement noise in the latter quantity. The correlation between the ERP and the co-skewness (computed with the US market, see Table \ref{ERP}) is found to be positive $\approx 0.36$, 
at variance with results on single stocks.} 
We have therefore extended our analysis to different contracts and different risk premia, and indeed find that in most cases, excess returns 
lead to a humped shape function $F_0(p)$, meaning that returns of large amplitude have a negative mean. This will lead us to propose a possibly universal definition of ``risk premium'' 
in terms of the skewness of the returns -- in a precise sense, the premium compensates for a tail risk that is systematically biased downwards. 

\section{Zooming into stock markets: the Fama-French factors}

We now consider the celebrated extended Fama-French model, which attributes excess stock returns to 4 components, corresponding to 4 orthogonal portfolios: the Equity Risk Premium discussed above (i.e. market minus the risk-free rate, MKT), SMB (``Small'' caps minus ``Big'' caps), UMD (``Up'' -- previous winners minus ``Down'' -- previous losers) and HML (High book to price -- ``value'' stocks minus Low book to price -- ``growth'' stocks), each of which is interpreted as a reward for a specific risk, even if the origin of said risk is not always clear.

The first, long market portfolio is tautologically not market neutral, while the other three are explicitly constructed in a market neutral way, in order to be orthogonal to the market risk factor MKT. 
SMB consists of being long the small cap stocks and short the large caps. This sounds like an intuitive risk premium strategy, since it seems plausible that small caps have a greater expected future growth but are more ``risky'' than large caps. However, if we look at Table \ref{USCAPvol}, we see that the volatility of large or small cap indices is practically the same in the US and that, if anything, volatility is actually {\it smaller} for small caps than for large caps in all other countries. Therefore, volatility alone is unable to explain the existence of a possible SMB risk premium. 

But if we now observe the above defined skewness $\zeta^*$ of large, mid and small cap indices reported in Table \ref{USCAPvol}, we clearly conclude that skewness does indeed become more negative as we move to smaller cap indices. This behaviour is furthermore quite universal: across the 10 cases we have investigated, only the Netherlands and New Zealand show a slight inversion of the skewness of mid and small cap stocks, while still maintaining the general 
trend noted above. Let us insist that we have also computed three other (more standard) estimates of the skewness with exactly the same conclusion. To add to the significance of this observation, we have built market neutral portfolios that are long 1\$ of the small-cap index, and short 1\$ of the large-cap index for each of the 10 countries we consider. We then globally risk manage these portfolios by rescaling the position with a 20-day moving average of their own volatility. In this way, we achieve approximately constant volatility throughout the simulation. For each country in our pool, the skewness of these portfolios is found to be negative, and they all yield positive performance, except one of them (Australia) -- see Table \ref{USCAPvol}, last column. This is quite significant, since these portfolios are completely decorrelated from one another.

\begin{table}
\begin{center}
\begin{tabular}{|c|c|c|c|c|c|}
\hline
\hline
Index & Start date & Cap & $\sigma$ (\%/day) & $\zeta^*$ & SR of SMB  \\
\hline
 &  & B & 1.1 & -0.66 & \\
S\&P & 1990 & M & 1.2 & -1.12 & 0.32 \\
 &  & S & 1.2 & -1.16 & \\
\hline
Russell & 1978 & B & 1.1 & -0.58 & 0.07 \\
Russell &  & S & 1.2 & -1.43 &  \\
\hline
 & & B & 1.3 & 0.08 &  \\
Japan & 1980 & M & 1.1 & -0.63 & 0.02 \\
 & & S & 1.0 & -0.94 &  \\
\hline
 & & B & 1.4 & -0.78 &  \\
France & 1999 & M & 1.1 & -1.66 & 0.40 \\
 & & S & 0.9 & -2.40 &  \\
\hline
  & & B & 1.4 & -0.92 &  \\
The Netherlands &  1995 & M & 1.3 & -1.71 & 0.33 \\
  & & S & 1.0 & -1.68 &  \\
\hline
 & & B & 1.2 & -1.68 &  \\
Italy & 2001 & M & 1.2 & -1.86 & 0.70 \\
 & & S & 1.0 & -2.98 &  \\
\hline
 & & B & 1.0 & -0.78 &  \\
Australia & 1994 & M & 1.0 & -0.88 & -0.27 \\
 & &  S & 1.0 & -1.41 &  \\
\hline
 & & B & 0.8 & -0.90 &  \\
New Zealand & 1998 & M & 0.7 & -1.23 & 0.39 \\
 & & S & 0.5 & -1.06 &  \\
\hline
 & & B & 1.7 & 0.15 &  \\
Poland & 1998 & M & 1.2 & -0.92 & 0.34 \\
 & & S & 1.2 & -1.17 &  \\
\hline
 & & L & 1.3 & -0.68 &  \\
South Africa & 1996 & M & 0.9 & -1.36 & 0.26 \\
 & & S & 0.7 & -1.59 &  \\
\hline
\hline
\end{tabular}
\caption{Volatility (in percent per day), and skewness of the daily returns of large(B)/mid(M)/small(S) cap indices in various countries. In the last
column, we give the Sharpe ratio of portfolios that are long small caps (S) and short large caps (B), and neutral on medium caps (M). Note that the volatility 
of small cap indices is not larger than that of large cap indices, while their skewness is subtantially more negative.}
\label{USCAPvol}
\end{center}
\end{table}

We can actually dwell further on this issue using the ``decile'' portfolios for the three Fama-French factors in the US equity market.\footnote{Available at {\tt http://mba.tuck.dartmouth.edu/pages/faculty/ken.french/}} 
We report in Table \ref{FFSMBvol2} the volatility, skewness and Sharpe ratio for being long each decile portfolio since 1950, for the three factors SMB, HML and UMD. Again, there is little evidence that the volatility 
plays any role at all for any of the 3 factors.
The skewness, on the other hand, becomes more negative for the portfolios with the highest returns for the SMB and UMD (momentum) factors. The smallest caps have (when traded together) a skewness $\zeta^* = -1.83$, 
whereas the largest caps have much 
weaker skewness $\zeta^* = -0.39$, confirming the conclusion above. A portfolio of winners has a skewness $\zeta^* = -0.90$ and is prone to large reversal events, whereas a portfolio of 
losers are essentially not skewed.\footnote{This may look surprising at first sight. In fact, the zero skewness of the ``Down'' portfolio comes from a compensation between the natural negative skewness of 
stocks, and a strong positive skewness contribution coming from periods where the market rebounds strongly after prolonged negative trends; see e.g. \citet{Kent}.} For HML, surprisingly, there is no such effect; if anything, the skewness appears to be slightly less negative for high deciles, i.e. for value stocks with large book to price ratios. Notice also that the correlation between skewness and Sharpe
ratios is clearly inverted for HML, see Table \ref{FFSMBvol2}. 

\begin{figure}
\begin{center}
\epsfig{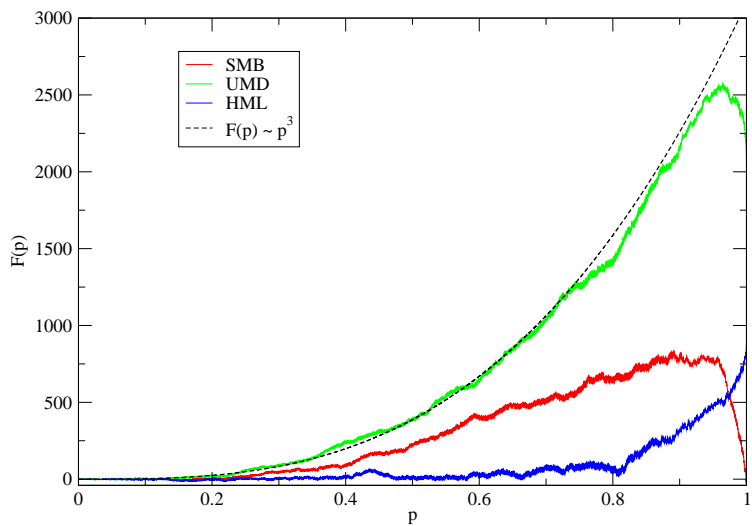}
\caption{The ranked amplitude P\&L $F(p)$ of the three Fama-French factors, SMB, UMD \& HML. Note the familiar humped shape for SMB and UMD, revealing strong negative skewness (respectively $\zeta^*=-1.39$ and $-1.38$), but an inverted behaviour for HML ($\zeta^*=+0.43$). The corresponding Sharpe ratios are, respectively, $0.14$, $0.67$ and $0.64$. We also show for reference the $p^3$ behaviour expected at small $p$ (for clarity, only shown for UMD).}
\label{skewFF}
\end{center}
\end{figure}

As a visual confirmation, we plot in Fig. \ref{skewFF} 
the ranked amplitude P\&L of the
three Fama-French factors in the US since 1950. Clearly, SMB and UMD show the typical humped shape observed for the market ERP in the previous section, therefore corresponding to a negative overall skewness equal, 
respectively, to $\zeta^*=-1.39$ and $-1.38$. For UMD, the 5\% largest events contribute to losses that amount to 20\% of the gains accumulated by the 95\% small to moderate returns, whereas for SMB nearly all the gains are erased by the 5\% largest 
events! For HML however the situation is inverted, reflecting the known anti-correlation between ``Value'' (HML) and ``Momentum'' (UMD): large returns contribute positively to the gains, therefore leading to a positive skewness $\zeta^*=+0.25$. We will discuss this feature of HML in Sect. 7-2.

\begin{table}
\begin{center}
\begin{tabular}{|c|c|c|c|c|c|c|c|c|c|c|}
\hline
\hline
 & 1 & 2 & 3 & 4 & 5 & 6 & 7 & 8 & 9 & 10 \\
\hline
SMB $\sigma$ (\%/d) & 0.83 & 1.00 & 1.00 & 0.98 & 0.97 & 0.92 & 0.93 & 0.94 & 0.93 & 0.96 \\
SMB $\zeta^*$ & -1.83 & -1.53 & -1.49 & -1.52 & -1.42 & -1.45 & -1.26 & -1.28 & -0.93 & -0.39 \\
SMB SR & 0.56 & 0.48 & 0.53 & 0.50 & 0.53 & 0.53 & 0.53 & 0.50 & 0.48 & 0.39 \\
\hline
UMD $\sigma$ (\%/d) & 1.48 & 1.21 & 1.05 & 0.99 & 0.95 & 0.93 & 0.92 & 0.94 & 1.01 & 1.23 \\
UMD $\zeta^*$ & 0.00 & -0.14 & 0.00 & -0.24 & -0.24 & -0.30 & -0.58 & -0.71 & -0.79 & -0.90 \\
UMD SR & -0.07 & 0.17 & 0.33 & 0.37 & 0.39 & 0.46 & 0.45 & 0.62 & 0.53 & 0.67 \\
\hline 
HML $\sigma$ (\%/d) & 1.05 & 0.97 & 0.93 & 0.95 & 0.94 & 0.92 & 0.91 & 0.97 & 0.98 & 1.10 \\
HML $\zeta^*$ & -0.70 & -0.69 & -0.52 & -0.68 & -0.52 & -0.66 & -0.59 & -0.65 & -0.55 & -0.36 \\
HML SR & 0.33 & 0.41 & 0.43 & 0.44 & 0.50 & 0.52 & 0.52 & 0.59 & 0.61 & 0.60 \\
\hline
\hline
LoV $\sigma$ (\%/d) & 1.9 & 1.5 & 1.3 & 1.2 & 1.1 & 1.1 & 1.0 & 0.9 & 0.9 & 0.7 \\
LoV $\zeta^*$ & -0.20 & -0.40 & -0.43 & -0.43 & -0.43 & -0.30 & -0.33 & -0.35 & -0.24 & -0.18 \\
LoV SR & 0.49 & 0.59 & 0.71 & 0.73 & 0.79 & 0.78 & 0.88 & 0.86 & 0.90 & 1.01 \\
\hline
\hline
\end{tabular}
\caption{Volatility, skewness and Sharpe ratio of the daily returns of the decile portfolios corresponding to SMB, UMD and HML in the US equity market since 1950. 
See also Fig. \ref{skewSR} for a plot of Sharpe ratios vs. skewness. Decile 1 corresponds, respectively, to 
Small caps, Down stocks and Low earning to price stocks, whereas decile 10 corresponds to Big caps, Up stocks and High book to price. For completeness, the last three lines corresponds to
the ``Low Vol'' (LoV) deciles (the first decile is the most volatile); see \citet{LowVol_us} for details.}
\label{FFSMBvol2}
\end{center}
\end{table}

Apart from the special HML case, the above findings give further credence to the idea that risk premium is actually 
{\it not} compensating investors for taking a symmetric risk (as measured by the volatility), 
but, as suggested by \citet{Kraus,Rietz,Barro,HarveySiddique} and \citet{SantaClara}, for carrying the risk of large losses.  
However, at variance with \citet{Rietz,Barro}, we do not attempt to define catastrophic macro-economic events and estimate their probability, 
but rather directly measure the skewness of these risk premia portfolios through our single humped function $F(p)$ and its integral $-\zeta^*$. 
These quantities immediately reveal the ``tail risk'' embedded in these strategies, i.e. that large returns systematically have a negative mean.

For completeness, we have also given in Table \ref{FFSMBvol2} the volatility, skewness and Sharpe ratios of the volatility deciles, 
clearly revealing the ``Low Vol'' anomaly, in line with \citet{Ang,Blitz,Ang2,LowVol,BAB,LowVol_us}. There is however no clear pattern in the 
skewness; in fact a market neutral ``Low Vol'' portfolio is found to have a slightly positive skewness (see Fig. \ref{skewSR} below).

\section{Risk Premia in the fixed-income world}

\begin{table}
\begin{center}
\begin{tabular}{|c|c|c|c|}
\hline
\hline
Grade & Yield (\%/y) & $\sigma$ (\%/y) & $\zeta^*$ \\
\hline
AAA & 4.7 & 5.1 & -0.87 \\
AA & 4.8 & 4.6 & -0.74 \\
A & 5.3 & 4.9 & -0.83 \\
BBB & 6 & 4.8 & -0.81 \\
BB & 7.7 & 4.0 & -1.78 \\
B & 9.5 & 4.8 & -1.99 \\
CCC and below & 15.5 & 7.3 & -1.94\\
\hline
\hline
\end{tabular}
\caption{Return and volatility (in percent), and skew of BofA US Corp. indices, using Global Financial Data ({\tt www.globalfinancialdata.com}). All data since 1997.}
\label{BOFAvol}
\end{center}
\end{table}

In the world of fixed income, credit risk is gauged by the investment grade of the corresponding bond, and one should in principle expect higher yields for lower investment grades (i.e. riskier bonds). 
We have studied corporate bond portfolios, using Global Financial Data time series for US bond total return data.\footnote{In this paper, we only study US corporate debt, but we believe our results have a more universal scope. For example, we have checked that a portfolio of international 10Y bonds also exhibits the same features.} 
We have used the BofA Corp. total return index suite, that provides daily bond returns since 1997 for grades going from AAA (safe) to CCC  and below (risky). We have summarised the average yield, volatility and skewness for each of these indices in Table \ref{BOFAvol}. The first striking observation is that although we cover a variety of grades and yields (ranging from $4.7 \%$ annual for AAA grade to $15.5 \%$ annual for CCC and below), the volatility of these 
indices is found to be remarkably constant at around $4.8 \%$ annualised, except for the lowest grade which has a $7.3 \%$ volatility. Similar to what we found above for stock indices corresponding to different market caps, the skewness is much more revealing than the volatility as it clearly drops between BBB and BB grades (see Table \ref{BOFAvol}).

\begin{figure}
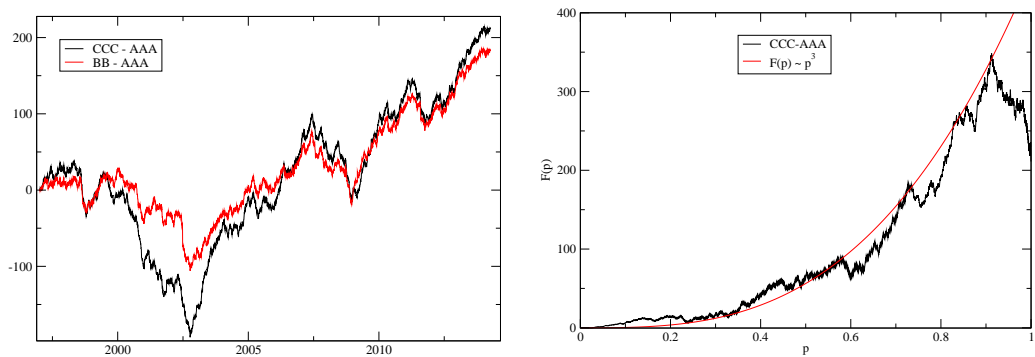

\begin{center}
\epsfig{file=pnlHYIG.eps, width=0.4\textwidth,angle=0}\hskip 0.5cm\epsfig{file=CCCAAA.eps, width=0.4\textwidth,angle=0}
\caption{Left: Performance of a risk-managed portfolio long a high yield index (CCC or BB), and short its AAA counterpart. SR: 0.54/0.45, t-stat: 2.2/1.9, $\zeta^*$:-0.84/-0.43. Right: Ranked amplitude P\&L of the
CCC/AAA portfolio, showing the characteristic humped shape of portfolios with downside tail risk.}
\label{pnlHYIG}
\end{center}
\end{figure}

Another way to elicit this skewness premium is to build a portfolio where we invest 1\$ every day in a high yield index (CCC or BB), and sell 1\$ of the AAA index. To achieve constant volatility throughout the simulation, the 
portfolio is (as above) risk-managed based on its past volatility, as measured by a 20-day moving average of the absolute returns of the strategy itself. The performance of the BB/AAA or CCC/AAA strategies is shown in Fig. \ref{pnlHYIG}, Left. The two strategies are similar and yield positive (but marginally significant) performance. While the explanation in terms of volatility difference is not very convincing, we again find that the ranked amplitude P\&L plot has a clear hump shape (see Fig. \ref{pnlHYIG}, Right), indicating again that large returns are systematically biased downwards. 

\section{Risk Premia in the FX world: the ``Carry Trade''}

Similarly to the previous section where we traded bonds of differing quality, one may want to buy short term debt in a country whose short term rate is high using 
borrowed money from a country whose short term rate is low. This idea is a textbook strategy in the world of FX finance called the ``Carry Trade''. 
In the absence of adverse exchange rate moves, this allows the investor to pocket the difference between the high interest rate and the low interest rate. 
This excess return is again usually thought of as a risk premium: one invests in the economy of a ``risky'' country, while getting financed in a safe country.
The questions we want to ask are, as above: a) is the excess return compensating for volatility? and b) is the strategy 
negatively skewed?

We have simulated this currency pair strategy on a pool of 20 developed countries using data from Global Financial Data ({\tt www.globalfinancialdata.com}), both for spot currencies and for interest rates, from 
Jan 1974 to Jan 2014.\footnote{The pool of countries is made up of: Australia, Canada, Switzerland, Czech Republic, Germany, Hungary, Israel, Iceland, Japan, Korea, Mexico, Norway, New Zealand, Poland, Sweden, Singapore, Thailand, UK, USA, South Africa.} We use intra-bank rates when they are available, and central bank discount rates otherwise. For each day, we consider all $20 \times 19/2$ currency ordered pairs, using the difference of interest rate 
in the two countries to order the pairs in a systematic way. We then make 10 deciles (following the Fama-French construction) corresponding to pairs with the smallest interest rate difference (i.e. the
smallest ``carry'') to the largest. We then compute, as above, the volatility, skewness and Sharpe ratio corresponding to these 10 portfolios.  The data is shown in Table \ref{FXC} and as small pluses in Fig. \ref{skewSR} below. We find again that skewness and Sharpe ratio are quite strongly correlated $\rho=-0.76$, with a correlation that is significantly larger in the period 1994-2014 than in the previous period (1974-1993). In this case however, we also find a substantial positive correlation between volatility and Sharpe ratio ($\rho=+0.78$). 
The Sharpe ratio of the fully diversified Carry strategy (i.e. with an equal weight over all ordered FX pairs) is ${\cal S} \approx 0.85$ for a skewness $\zeta^* \approx -0.94$. 

The above results contrast somewhat with the recent conclusion of \citet{Jurek} who finds that the profitability of the FX carry trade is 
not substantially degraded when hedged with out-of-the-money currency options. This leads him to the conclusion that the carry trade P\&L is {\it not} related to the (implied) tail-risk. However, we have not been able to confirm Jurek's findings. We found that when hedging the carry trade strategy with at-the-money options instead of out-of-the-money options (for which our data is less reliable), the carry P\&L instead completely vanishes. Since out-of-the-money volatilities are usually more expensive than at-the-money volatilities, we conclude that
when hedged with out-of-the-money currency options, the carry trade P\&L would become negative, in agreement with our general skewness story.

\begin{table}
\begin{center}
\begin{tabular}{|c|c|c|c|c|c|c|c|c|c|c|}
\hline
\hline
 & 1 & 2 & 3 & 4 & 5 & 6 & 7 & 8 & 9 & 10 \\
\hline
FX Carry $\sigma$ (\%/d) & 0.45 & 0.47 & 0.45 & 0.50 & 0.55 & 0.58 & 0.64 & 0.68 & 0.74 & 0.93 \\
FX Carry $\zeta^*$ & -0.41 & -0.35 & -0.60 & -0.50 & -0.87 & -0.97 & -0.86 & -1.04 & -0.81 & -1.05 \\
FX Carry SR & -0.28 & -0.05 & 0.56 & 0.44 & 0.51 & 0.40 & 0.78 & 0.62 & 0.91 & 1.04 \\
\hline
\hline
\end{tabular}
\caption{Volatility, skewness and Sharpe ratio of the daily returns of the decile portfolios corresponding to FX Carry on G20 currencies since 1974. 
See also Fig. \ref{skewSR} for a plot of Sharpe ratios vs. skewness. Decile 1 corresponds to pairs with small interest rate difference whereas decile 10 corresponds to pairs with small interest rate difference. Note that in this case, there is a positive correlation between the Sharpe ratio and the volatility, on top of the 
now familiar negative correlation between Sharpe ratio and skewness.}
\label{FXC}
\end{center}
\end{table}

\section{A paradigmatic example: Risk Premia in option markets}

The textbook example of a risk premium strategy is selling options, i.e. insurance contracts against large movements of financial assets -- for example stock indices. In this case, there is a built-in asymmetry 
between the risk taken by the option seller (the insurer) and the option buyer. The option seller bears the risk of arbitrarily large losses, and hence the skewness of the P\&L of the insurer is (almost by
definition) negatively skewed -- even after hedging in a world where perfect replication is not possible \citep[see the discussion in][in particular Ch.\ 14]{Book}. Conversely, the skewness of the option buyer is positive. Therefore, one should expect that option sellers require a ``skew premium'' as a compensation for the asymmetric risk they 
agree to cover. In order to confirm this rather basic intuition, we have computed the performance of selling variance swaps, which are a basket of options 
whose return is directly proportional to the difference between realised and implied variance. This difference is exactly the risk premium we seek to elicit, so these baskets are perfectly suited to our purpose. 

\begin{figure}
\begin{center}
\epsfig{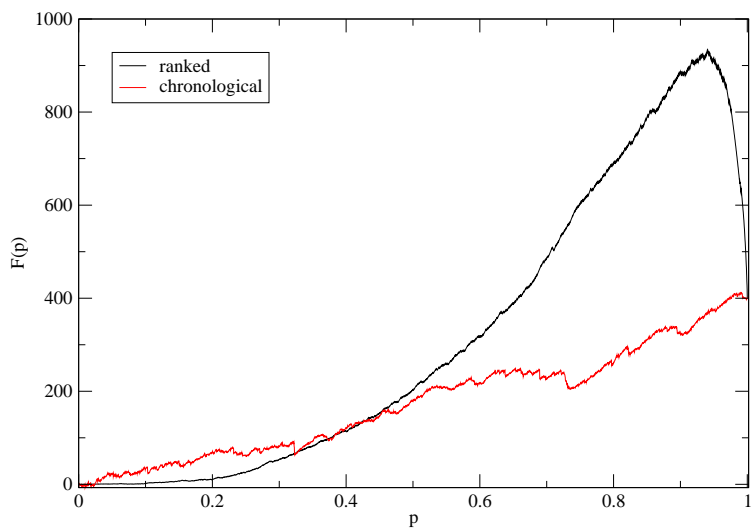}
\caption{Performance of a risk managed short-option strategy, equi-weighted on S\&P 500, DAX, Crude oil, Gold, EURXUSD and GBPXUSD variance swaps from Jan 1996 until Jan 2014. The Sharpe ratio is $1.26$. We also show the ranked P\&L which reveals the strongly 
skewed nature of selling volatility. }
\label{pnlVSW}
\end{center}
\end{figure}

We test this risk premium strategy on a variety of products, since the above skew premium argument should be a universal feature valid across the board, from commodity (Crude Oil, Gold) and currency options (EURXUSD, GBPXUSD) to stock index variance swaps (S\&P 500, DAX) since Jan 1996. 
Selling each of these variance swap contracts, and risk-managing using the last 20-day's volatility of the P\&L itself (as we did in the previous sections), leads to the global performance shown in Fig. \ref{pnlVSW}. As we can see, the return is clearly positive, with a t-stat over 4, while the strategy is strikingly skewed, as expected for this paradigmatic risk premium strategy. We have summarised the Sharpe ratio and the skewness $\zeta^*$ for all six contracts in Table \ref{skewVSW}. Clearly the skewness is indeed negative in all cases. In fact, the 
skewness $\zeta^*$ is substantially larger than all the examples considered so far in this paper and the performance of the individual short var-swap strategies is also somewhat better than all the risk premia
analysed above. Interestingly, shorting the VIX future instead leads to a P\&L that is both less skewed and with a lower Sharpe ratio (see Fig.  \ref{skewSR} below). This again suggests that there might exist a systematic relation between risk premia and skewness. This is the point we 
now discuss. 

\begin{table}
\begin{center}
\begin{tabular}{|c|c|c|}
\hline
\hline
Underlying & Sharpe Ratio  & $\zeta^*$ \\
\hline
S\&P 500 & 1.47 & -4.64 \\
DAX & 0.79 & -5.53 \\
Crude oil & 0.74 & -5.28 \\
Gold & 0.71 & -4.79 \\
EURXUSD & 0.88 & -5.32 \\
GBPXUSD & 0.38 & -5.01 \\
\hline
Total & 1.26 & -4.61 \\
\hline
\hline
\end{tabular}
\caption{Sharpe ratio and skewness $\zeta^*$ of short var-swap, risk managed strategies on different underlyings since Jan 1996.}
\label{skewVSW}
\end{center}
\end{table}

\section{Risk premium is Skewness premium}

\subsection{A remarkable plot}

The consistent picture that seems to emerge from all the above empirical results is that ``risk premia'' are in fact very weakly (if at all) related to the volatility of an investment, but rather to its skewness, or
more precisely to the fact that the largest returns of that investment are strongly biased downwards. This has already been discussed several times in the context of stock markets, from different standpoints (economic, 
econometric and financial). For example, the ``Low Volatility'' puzzle goes against the intuitive argument that more (volatility) risk must be accompanied by higher 
average returns. Similarly, we find no evidence of a higher volatility in the small caps, large momentum, or high value stocks that would explain the excess returns associated to these factors, while skewness seems to 
play an important role. Perhaps obviously, the risk premium collected by shorting
options appears to be compensating for the substantial skewness of an insurance P\&L. Table \ref{Table_summ} summarises the correlation between Sharpe ratios, volatility and skewness for all the strategies discussed above. 

\begin{table}
\begin{center}
\begin{tabular}{|c|c|c|}
\hline
\hline
Underlying & Vol/SR corr.  & Skewness/SR corr \\
\hline
Bonds & -0.69 & -0.36 \\
Intl. IDX & -0.45 & -0.38 \\
SMB & -0.42 & -0.89 \\
UMD & -0.63 & -0.85 \\
FX Carry & +0.78 & -0.76 \\
\hline
HML & +0.03 & +0.64 \\
LoV & -0.98 &  +0.23 \\
TREND & +0.23 & +0.58 \\
\hline
\hline
\end{tabular}
\caption{Correlation coefficient $\rho$ between volatility and Sharpe ratio, and between skewness and Sharpe ratio for all
strategies investigated in this paper, and for a 50-day trend following strategy across various futures markets. 
In most cases, and counter-intuitively, correlation with volatility is found to be {\it negative} or zero -- note in particular
the Low Vol anomaly (LoV). Correlation with
skew, on the other hand, is chiefly negative, showing that the 
main determinant of risk premium must be skewness and not volatility. Trend, HML and LoV are clear outliers.}
\label{Table_summ}
\end{center}
\end{table}

\begin{figure}
\begin{center}
\epsfig{file=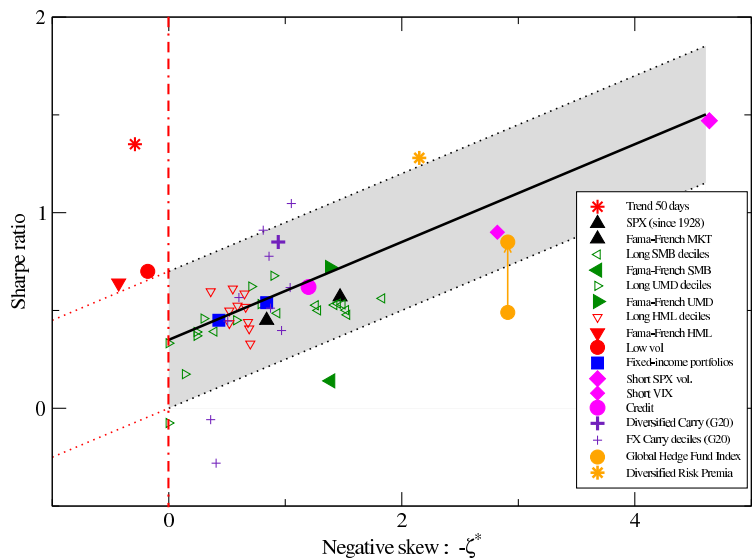, width=0.6\textwidth,angle=0}
\caption{Sharpe ratio vs Skewness $-\zeta^*$ of all assets and/or strategies considered in this paper (see legend). 
 The relatively low Sharpe of the HFRX index can be explained by the hedge fund fees (the
vertical arrow corresponds to a tentative estimate of these fees). The plain line ${\cal S} \approx 1/3 - \zeta^*/4$ corresponds to the regression 
line through all risk premia marked with filled symbols (excluding trend following). The two dotted lines correspond to 
a 2-$\sigma$ channel, computed with the errors on the SR and on the skewness of the Fama-French strategies. The small empty 
triangles correspond to the Fama-French decile portfolios.
Trend following is a clear outlier marked as a red point, characterised by a positive 
skewness {\it and} a positive SR. The same is true, perhaps to a lesser extent, of HML. Note also the interesting difference 
between short var-swap and short VIX strategies. }
\label{skewSR}
\end{center}
\end{figure}

In order to reveal a possibly universal relation between excess returns and skewness, we have summarised all our results above in a single scatter plot, Fig. \ref{skewSR}, where we show the Sharpe Ratios of different 
portfolios/strategies as a function of their (negative) skewness $-\zeta^*$. Quite remarkably, {\it all but one} fall roughly on the regression line ${\cal S} \approx 1/3-\zeta^*/4$. This is the central result of our paper. The two parallel dashed lines correspond to a 2-$\sigma$ channel, computed with the errors on the SR and the skewness of the Fama-French strategies (other strategies have even larger error bars), see Table \ref{STDsrskew}. The glaring
exception, that we discuss in more detail below, is the 50 day trend following strategy on a diversified set of futures contract since 1960 \citep[see][for details]{Trend_us}. The other outliers, although less clear-cut, 
are the Fama-French HML factor and the Low-Vol (LoV) strategy, which also have positive skewnesses.

Interestingly, we have also considered the returns of a portfolio of four credit indices since 2004, and of the HFRX global hedge fund index, which provides us 
with daily data since 2003. Although this history is relatively short, the Sharpe ratio/skewness of both credit and of the hedge fund index fall 
in line with the global behaviour (once fees are taken into account in the case of the HFRX \footnote{The Sharpe ratio of HFRX with fees is $0.48$, and we have corrected this to yield a Sharpe ratio without fees of $0.85$. This assumes that funds run at a volatility of around $\sim 10 \%$ and have a classic 2 and 20 fee
structure.}).  Let us insist once more that Fig. \ref{skewSR} is a suggestive summary of our empirical result and not the prediction of 
a valuation theory. In this respect, Fig. \ref{skewSR} looks qualitatively similar for other low-moment 
definitions of skewness, but completely loses its structure when the skewness is replaced with a measure of coskewness with the equity market, as a naive extension of the CAPM model, along the lines of \citet{Kraus,HarveySiddique} would suggest.

\begin{table}
\begin{center}
\begin{tabular}{|c|c|c|}
\hline
Strategy & Error on the skewness & Error on the SR \\
\hline
\hline
FXR pairs & 0.19 & 0.18 \\
Fixed income & 0.27 & 0.24 \\
Agg. varswaps & 0.22 & 0.24 \\
Global hedge fund index & 0.41 & 0.30 \\
Trend 50d fut.  & 0.16 & 0.14 \\
Fama-French MKT & 0.12 & 0.13 \\
Fama-French SMB & 0.13 & 0.13 \\
Fama-French HML & 0.12 & 0.13 \\
Fama-French UMD & 0.12 & 0.13 \\
\hline
\hline
\end{tabular}
\caption{Error bars on the SR and skewness of the strategies appearing in Fig. \ref{skewSR}. 
The errors on the skewness are estimated using a standard bootstrap method.}
\label{STDsrskew}
\end{center}
\end{table}

Fig. \ref{skewSR} not only efficiently summarises all our results, but also suggests both a classification and a grading of different investment strategies. It is tempting to {\it define} a risk premium strategy 
as one that compensates for the skewness of the returns, in the sense that its Sharpe ratio lies close to the ``skew-rewarding'' line ${\cal S} \approx 1/3-\zeta^*/4$. 
Interestingly, the aggregate returns of hedge fund strategies fall on this line, suggesting
that a substantial fraction of hedge fund strategies are indeed risk premia.

Strategies that lie significantly below this line take too much tail risk for the amount of excess return. Contrarily, strategies that lie significantly above this line, in particular those with {\it positive} skewness such as trend following, seem to get the best of both 
worlds. But by the same token, these strategies cannot meaningfully be classified as risk premia;\footnote{Barring, of course, the existence of an unidentified risk factor that has never materialised, but that 
would contribute to a negative skewness if it did.} 
rather, as argued in \citet{Trend_us}, these excess returns  must represent genuine market anomalies, or ``pure $\alpha$''. 

\subsection{Three interesting exceptions: HML, LoV \& Trend following}

Let us dwell a little longer on our outliers: trend following, HML and LoV. The daily skewness of these portfolios is significantly positive, respectively 
$\zeta^*=+0.43$, $\zeta^*=+0.27$ and $\zeta^*=+0.18$, and yet their performance is highly significant and universal across geographical zones. 

The concept of risk premium applied to HML and to LoV is indeed problematic. Intuitively, high book-to-price corresponds to ``value'' stocks that are usually considered 
safe and defensive, while low book-to-price corresponds to ``growth'' stocks, i.e. a risky bet on future earnings. This is confirmed by the 
results of Table \ref{FFSMBvol2} above, which shows that growth stocks are indeed more negatively skewed than value stocks, in line with our 
definition of risk, but indeed making HML an outlier in Fig \ref{skewSR}.\footnote{However, to add to the confusion, the
skewness of the monthly returns of the HML strategy, although more noisy, appears to change sign ($\zeta^*_{\text{month}} = - 0.56$). 
This suggests that more work is necessary to fully understand the nature of the HML factor, which is actually known to have long periods of draw-downs 
(like during the Internet bubble) where the effect appears to be more in line with a standard risk premium interpretation.} For LoV, the
paradox is even greater: in what sense can it be risky to invest on low-volatility stocks and short high-volatility stocks? It seems to us quite
reasonable that LoV cannot be understood as a risk premium.

Turning now to the nature of trend following, it could in fact have been anticipated that its skewness should be positive. 
This is because trend following is 
profitable, by definition, when the long term realised volatility is larger than the short term volatility \citep{Trend_gamma}. 
In other words, trend following is akin to a ``long gamma'' strategy, and is thus expected to have an oppositely signed skewness to that of options. 
Correspondingly, the skewness of monthly returns is found to be 
even more positive than that of daily returns ($\zeta^*_{\text{month}} = + 1.72$). Hence, the highly significant excess returns of trend following strategies in the 
last two centuries  \citep[see][for a recent discussion]{Trend_us} seem difficult to ascribe to any reasonably defined risk premium. The profitability of trend following 
appears to be a genuine market anomaly, plausibly of behavioural origin.\footnote{The attentive reader 
might be puzzled by the fact that trend following has a positive skewness, while market neutral momentum (i.e. UMD) has a negative skewness, as shown above. Still, the two
strategies should hinge upon the same underlying behavioural biases. This paradox will be discussed in a forthcoming publication.}
We find our plot Fig. \ref{skewSR} interesting in the sense that trend following and (albeit to a lesser extent) HML and LoV can be clearly identified as outliers.

\section{Discussion and Conclusions} 

\subsection{Skewness vs. co-skewness}

Our central result, conveyed by Fig. \ref{skewSR} above, is that skewness is indeed the main determinant of risk premia, with an approximately 
linear relation between the Sharpe ratio ${\cal S}$ of a risk premium strategy and its skewness $\zeta^*$. 

From a formal point of view, this finding can be qualitatively interpreted within the classical framework of utility theory. Provided the third derivative of the 
utility function is positive, skewness-comparable P\&Ls can be ordered, and negatively skewed strategies should be compensated by higher returns to remain
attractive, as amply confirmed by lottery experiments \citep{Lotteries1,Lotteries2}. However, previous attempts to include formally the effect of skewness in 
valuation theories have led to an extended formulation of the CAPM \citep{Kraus,HarveySiddique}, where the {\it co-skewness} of a given stock with the market -- rather than
the skewness itself -- should determine the excess return of that stock. The idea here is that idiosyncratic skewness is diversifiable and should not lead to excess 
returns; only the exposure to global market shocks should play a role. As shown by \citet{Kraus,HarveySiddique,Harvey} \citep[and more recently in][]{Kelly}, this idea seems to have 
clear merits in the context of stock markets.

However, this is not the route we have taken in this paper, for several reasons. First, we tend to be very suspicious about equilibrium arguments leading to CAPM-like specifications. 
As amply demonstrated empirically -- including in the present study -- market anomalies are numerous and strong, and theoretical predictions based on arbitrage and rational 
behaviour not very compelling. Second, our results do {\it not} concern the behaviour of individual stocks but rather focus on the profitability of ``factors'' in a wide sense --
from Fama-French factors to bond, FX and volatility carry trades. Transposing the coskewness idea of \citet{Kraus,HarveySiddique} in the present context would first require 
the definition of a global ``risk factor'' that drives all risk premia, much as the market factor drives individual stocks. This is essentially the
content of the proposal of \citet{Rietz,Barro,SantaClara}: all these risk premia in fact represent exposure to the same ``catastrophic'' Black-Swan risk, that would -- if realised -- spread out over many different 
asset classes. We have tried to test this idea directly by studying the correlation matrix of all the risk premia strategies studied above, with the hope of identifying a dominant 
mode in the PCA, that would define a global risk factor. Perhaps surprisingly, we found that the top and second largest eigenvalue of this correlation matrix are not clearly 
separated (at variance with the correlation matrix of single stocks, that lead to a clear separation between the market and other sub-dominant factors). Furthermore, the 
structure of the top eigenvector is not stable in time, which means that it cannot be used as a benchmark to define a meaningful co-skewness. When (arbitrarily) defining the SPX as 
the global risk factor, the risk premia co-skewnesses defined {\`a la} \citet{Kraus,HarveySiddique} were found to be uncorrelated with the corresponding Sharpe ratio, i.e. all the
structure suggested in Fig. \ref{skewSR} disappeared. When testing the co-skewness idea restricted to international equity markets, we found (see Table \ref{ERP} above) that the Sharpe ratio of the ERP
is actually positively (rather than negatively) correlated with the co-skewness with the US market.

\subsection{Skewness, premia \& crowded trades}

\begin{figure}
\begin{center}
\epsfig{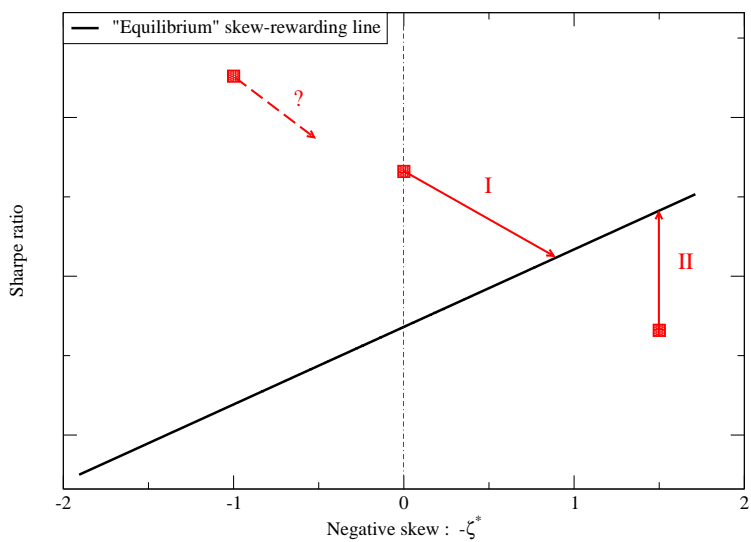}
\caption{
The skew-reward trade-off line ${\cal S} \approx 1/3-\zeta^*/4$ and different risk premium dynamical paths. A first path (I) starts with a symmetric asset that pays dividends, and therefore attracts investors. 
The trade becomes crowded and this creates potential de-leveraging spirals and downside tail risk. A second path (II) corresponds to an initially skewed payoff (like options) that moves up in Sharpe ratio as a result of skewness aversion and 
a dearth of option sellers. The path indicated with a question mark (?) might describe the fate of pure alpha strategies as they become more crowded and prone to deleveraging spirals of the type discussed in \citet{BP}.}
\label{skewMECH}
\end{center}
\end{figure}

Instead of trying to develop a formal equilibrium argument to explain our findings, we now propose a hand-waving intuitive picture of the mechanism 
enforcing the trade-off between skewness and excess returns. Our view is that there may in fact be {\it two fundamentally different} ways in which risk premia 
reach the skew-reward trade-off line in Fig. \ref{skewSR}: 
\begin{itemize}
\item In many situations, one 
buys an asset because of an expected stream of payments, like dividends for stocks, coupons for bonds, interest rate differences for currency pairs, etc. 
This intrinsic source of returns attracts a crowd of investors that generate a price increase 
but simultaneously create the risk of a crash, induced by a self-fulfilling panic or ``bank run'' mechanism, due to the crowdedness of the trade. Schematically:
\be \nonumber
{\rm dividends} \to {\rm buyers} \to {\rm crowdedness} \to {\rm price~increase~and~downside~tail~risk}.
\ee
As illustrated in Fig. \ref{skewMECH}, crowdedness decreases returns and increases downside tail risk. Both effects limit the number of additional investors and stabilise the market 
around an acceptable skewness/excess return trade-off. 
\item In other situations, downside tail risks pre-exist and it is their very existence that leads to excess returns. A perfect example is provided by option markets. 
Since options are insurance contracts, their payoff profile is {\it by construction} skewed - negatively for option sellers and positively for option buyers. 
In an efficient market, the fair price of options is such that their average payoff is zero, no risk premium exists and the Sharpe ratio of being long or short options is zero. However, the presence of fat tails and a natural investor preference for positive skewness leads to a rise in the market price of options as investors seek insurance against sudden adverse price movements. The negative skewness from the short option position, arising from the rare large payouts remains constant and ever present. The Sharpe ratio of the short option position in an inefficient market  rises to reflect the extra premium required to act as insurer. We would therefore summarise this scenario as:
\be \nonumber
{\rm downside~tail~risk} \to {\rm dearth~of~option~sellers} \to {\rm increased~premium~and~Sharpe},
\ee
until the risk premium collected by selling options becomes large enough to lure in enough supply to cap the price of these options.
\end{itemize}

A crucial assumption for the first mechanism to work is that negative skewness is induced by crowdedness and de-leveraging spirals, as was convincingly argued in \citet{BP}.
While it seems to us highly plausible that this is the case, it would be satisfying to find some direct empirical evidence confirming the above ``crowdedness'' story. 
Interestingly, we have checked that the 50-day trend following strategy, that has attracted a large amount of assets 
in the last 20 years, is characterised by a positive skewness, as shown in Fig. \ref{skewSR}, that has nevertheless been slowly drifting towards zero in recent years. Similarly, the skewness of 
the carry trade has significantly increased (i.e $\zeta^*$ has become more negative) in the period 1994-2014 as compared to the period 1974-1994, perhaps indicating  that the carry trade is more popular than it used to be. 

\subsection{Diversifying skewness?}

From a practical point of view, our results provide an objective definition of risk premia strategies and a criterion to assess their quality
based on the comparison 
between Sharpe ratio and  skewness, suggested by Fig. \ref{skewSR}. Clearly, not all excess returns can be classified as risk premia -- trend following being one glaring counter-example. On the other hand, if 
the different risk premia are somewhat decorrelated, as suggested by the absence of a global ``risk mode'' in the correlation matrix discussed above, 
then skewness can be diversified away, allowing risk premia portfolios to move {\it above} the regression line shown in Fig. \ref{skewSR}. That this may indeed be possible 
is illustrated by the orange ``star'' shown in Fig. \ref{skewSR}, 
corresponding to a synthetic Diversified Risk Premia portfolio, with an equal weight on Long stock indices, Short Vol, FX Carry and CDS indices. 
We believe that this is what good ``alternative beta'' managers should strive to achieve.

\section*{Appendix: Ranked Amplitude P\&Ls and an alternative definition of Skewness}

Let us consider a random variable $r$ (the returns) with a certain probability density $P(r)$. We assume that $r$ has been standardised (i.e. $r$ has zero mean and unit variance). We will denote as $x = |r| \geq 0$ the 
amplitude of $r$. From the definition of the ranked P\&L function $F_0(p)$, one has: 
\be
F_0(p) = \int_0^{x(p)} {\rm d}y \,  y \left[P(y) - P(-y)\right],
\ee
where $x(p)$ is the p-quantile of $|r|$, defined as $p = \int_0^{x(p)} dr (P(r) + P(-r))$. We will introduce the symmetric and antisymmetric contributions to $P$ as:
\be
P_s(r) = P(r) + P(-r), \qquad P_a(r) = P(r) - P(-r).
\ee
Note that for generic distributions, $P(r) \approx_{r \to 0} P(0) - P'(0) r + ...$, leading to lowest order (when $p \to 0$) to $x(p) \approx p/2P(0)$ and therefore 
\be
F_0(p) \sim -P'(0)/12 \times (p/P(0))^3,
\ee
i.e. a generic $\propto p^3$ behaviour for small $p$. 

Now, our definition of skewness is:
\be
\zeta^* := - 100 \int_0^1 {\rm d}p \, F_0(p) = -100 \int_0^\infty {\rm d}x \, P_s(x) \int_0^{x} {\rm d}y \, y P_a(y).
\ee
It is interesting to give an alternative, intuitive interpretation of this definition. After simple manipulations, one finds:
\be
F_0(p) = E [|r| | r < -x(p)] - E [|r| | r > -x(p)].
\ee
$F_0(p)$ therefore compares the average amplitude of large negative and large positive returns. $\zeta^*$ is an average of this difference over 
all possible quantile choices. It might also be useful to relate $\zeta^*$ to the standard definition of skewness $\zeta_3$ (defined through the third cumulant of a distribution) in the limit of weakly non-Gaussian distributions:
\be
P(r) =  \left[ 1 - \frac{\zeta_3}{3 !} \frac{d^3}{dr^3} + \frac{\kappa}{4 !} \frac{d^4}{dr^4} + ...\right] \frac{1}{\sqrt{2\pi}} e^{-r^2/2},
\ee
where $\kappa$ is the kurtosis; finally leading to 
\be
\zeta^* = \frac{25}{6 \pi} \zeta_3 \,( 1 - \frac{\kappa}{24} +...) \approx 1.273 \, \zeta_3 \, ( 1 - \frac{\kappa}{24} + ...)
\ee
Note that $\zeta^*$ does not require the existence of the third moment of the distribution, and remains well defined for all distributions with a finite first moment. In fact, an even weaker condition is sufficient: the distribution should fall out faster than $|r|^{-3/2}$. 

\begin{figure}
\begin{center}
\epsfig{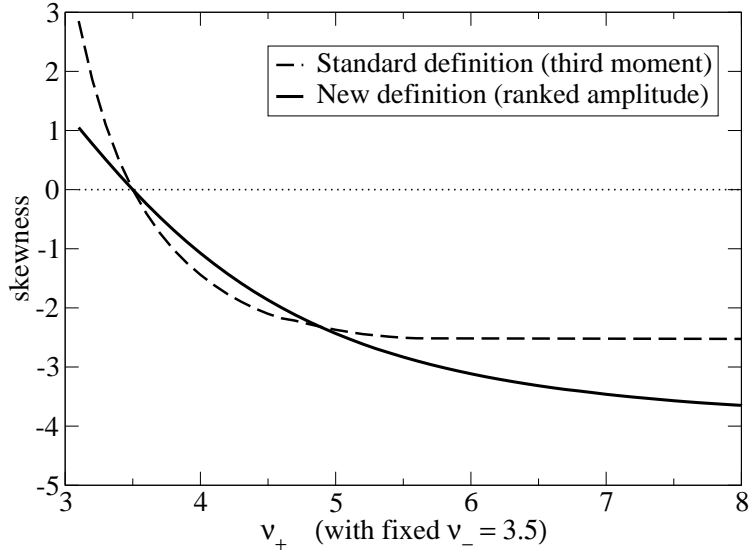}
\caption{$\zeta_3$ and $\zeta^*$ as a function of $\nu_+$ for a fixed value of $\nu_-=3.5$, for the assymetric Student-t
distribution defined in the text.
}
\label{skew-theory}
\end{center}
\end{figure}

It is interesting to compare $\zeta^*$ to the classical definition of skewness $\zeta_3$ is a concrete case. We choose an assymetric Student-t
distribution, defined as \citep{Jones}:
\be
P(r) = {\cal N} \left(1 + \frac{r}{\sqrt{\frac{\nu_++\nu_-}{2} + r^2}}\right)^{\frac{\nu_-+1}{2}}\left(1 - \frac{r}{\sqrt{\frac{\nu_++\nu_-}{2} + r^2}}\right)^{\frac{\nu_++1}{2}}
\ee
which behaves asymptotically as:
\be
P(r) \sim_{r \to \pm \infty} {\text{const.}} |r|^{-1-\nu_{\pm}}.
\ee
The classical skewness $\zeta_3$ is finite only when $\nu_\pm > 3$, whereas $\zeta^*$ remains finite as long as $\nu_\pm > 1/2$. While 
the latter condition is always satisfied by financial data, many authors have reported that $\nu_\pm$ is actually close to $3$ for most
markets. As a numerical exercice, we compute both $\zeta_3$ and $\zeta^*$ as a function of $\nu_+ > 3$, for a fixed value of $\nu_- = 3.5$.
The results are shown in Fig. \ref{skew-theory}. Clearly, the skewness of the distribution is positive when $\nu_+ < \nu_-$ and negative otherwise. What we see is that, as expected from the above general formula, $\zeta^*$ and $\zeta_3$ behave similarly when they are both small. 
However, as $\nu_+$ decreases towards $3$, $\zeta_3$ diverges whereas $\zeta^*$ remains well behaved. In the opposite direction, we see 
that $\zeta_3$ quickly saturates as $\nu_+$ increases, while $\zeta^*$ continues to decrease. Therefore, $\zeta^*$ is a better discriminant of the assymetry of the distribution.

Finally, let us assume that $F_0(p)$ is a humped function of $p$ with a single maximum, corresponding to a negatively skewed distribution. This means that necessarily $P_a(y>0)$ is negative for large enough $y > y^*$ and 
positive for smaller $y$s, and vice-versa for $y <0$. Now the cumulative function of $P$ compared to its symmetrised version is precisely the cumulative of $P_a(y)$:
\be
G(y)=\int_{-\infty}^y {\rm d}r [P(r) -P_s(r)] = \int_{-\infty}^y {\rm d}r P_a(r)
\ee
Since $P_a(r)$ vanishes three times and is positive for large negative $r$, it is clear that $G(y)$ has two symmetric maxima located at $\pm y^*$ and a minimum for $y=0$.  Now using $0= F_0(p=1) < y^* \int_{0}^{\infty} {\rm d}r P_a(r)$ and $G(0) = - \int_{0}^{\infty} {\rm d}r P_a(r)$, one immediately finds that $G(0) < 0$ for non degenerate distributions. This proves that $G(y)$ crosses zero twice and only twice, and therefore that $P(r)$ and $P_s(r)$ are 
skewness-comparable in the sense of \citet{Oja} whenever $F_0(p)$ has a unique maximum (or minimum).

\end{document}